\documentclass[prb,aps,twocolumn,amsmath]{revtex4}

\usepackage{graphicx}
\usepackage{epsfig}
\begin{document}
\title{Spin relaxation in (110) and (001) InAs/GaSb superlattices}
\author{K. C. Hall, K. G\"{u}ndo\u{g}du, E. Altunkaya, W. H. Lau, Michael E. Flatt\'{e}, and Thomas F. Boggess}
\affiliation{Department of Physics and Astronomy and Optical Science and Technology Center, The University of Iowa, Iowa City,\\
Iowa 52242}
\author{J. J. Zinck, W. B. Barvosa-Carter, and S. L. Skeith}
\affiliation{HRL Laboratories, LLC, 3011 Malibu Canyon Rd., Malibu, California 90265}
\begin{abstract}
We report an enhancement of the electron spin relaxation time
($T_1$) in a (110) InAs/GaSb superlattice by more than an order of
magnitude (25 times) relative to the corresponding (001)
structure.  The spin dynamics were measured using polarization
sensitive pump probe techniques and a mid-infrared, subpicosecond
PPLN OPO.  Longer $T_1$ times in (110) superlattices are
attributed to the suppression of the native interface asymmetry
and bulk inversion asymmetry contributions to the precessional
D'yakonov Perel spin relaxation process.  Calculations using a
nonperturbative 14-band nanostructure model give good agreement
with experiment and indicate that possible structural inversion
asymmetry contributions to $T_{1}$ associated with compositional
mixing at the superlattice interfaces may limit the observed spin
lifetime in (110) superlattices. Our findings have implications
for potential spintronics applications using InAs/GaSb
heterostructures.
\end{abstract}

\maketitle

Semiconductor heterostructures based on the InAs/GaSb/AlSb system
have gained attention recently for potential applications in the
rapidly growing field of semiconductor-based
spintronics,\cite{Cartoixa:2001,Hammar:2002,Gilbert:2002,Heida:1998}
in which the focus is to develop novel electronic devices and
circuits that exploit the spin property of the
electron.\cite{Wolf:2001,ALS:book} By utilizing some of the unique
characteristics of the InAs/GaSb/AlSb system, including the strong
spin orbit interaction in InAs and GaSb, the ability to form a
variety of interface types (type-I, type-II staggered, and type-II
broken gap), and the high electron mobility of InAs, it may be
possible to realize a host of novel high-speed spin-sensitive
electronic devices.  A crucial factor in the design of any
spin-sensitive electronic device is the electron spin relaxation
time (T$_{1}$), which must be sufficiently long to allow for
transport and$/$or processing of the spin-polarized electrons.  In
InAs/GaSb/AlSb heterostructures, which contain bonds between
semiconductor constituents with no atom in common, the nature of
the interfaces can have a dramatic influence on the electron spin
dynamics.  The electron spin relaxation time was recently measured
in (001) InAs/GaSb superlattices,\cite{Olesberg:2001} where it was
found that the asymmetric potential at the interfaces, referred to
as native interface asymmetry
(NIA),\cite{ALS:book,Olesberg:2001,Lau:2002,Vervoort:1999,Krebs:1996}
strongly dominates the precessional spin relaxation process.  In
addition to affecting the band gap and intervalence band
absorption properties of these superlattices,\cite{Lau:2002} NIA
leads to spin relaxation times $\leq$ 1 ps, which are more than an
order of magnitude shorter than expected considering bulk
inversion asymmetry (BIA)\cite{Lau:2001} contributions alone.

\begin{figure}[htb]\vspace{20pt} 
    \includegraphics[width=7.0cm]{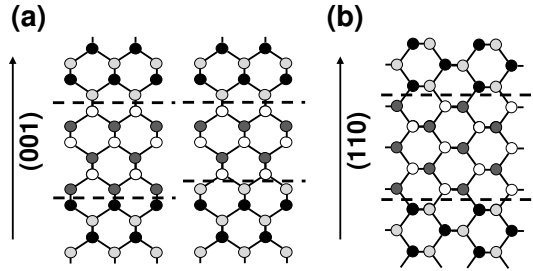}
    \caption{Schematic diagram of the interfaces in InAs/GaSb superlattices for (a): (001) and (b): (110) directions of growth.
      The shades indicate, in order of increasing darkness, Sb, In, Ga, As.  Since InAs and GaSb have no atom in common, for (001) superlattices the
      interface potential is intrinsically asymmetric, referred to
      as NIA.  The two types of bonding configurations for (001)
      superlattices are shown: different bond composition at the
      two interfaces (InSb and GaAs), or bonds of the same
      composition (InSb is shown) but different orientation.  In
      contrast, for (110) no-common-atom superlattices, NIA is absent because the interfaces are mixed, containing equal
numbers of both anions and cations, leading to a symmetric
interface potential.}
    \label{fig:Interfaces}
\end{figure}

The interface contribution to spin relaxation in these
no-common-atom superlattices may be removed through (110) growth,
since the mixed anion-cation interface planes in this orientation
lead to a symmetric interface potential. (See
Fig.~\ref{fig:Interfaces}.) Since NIA strongly dominates spin
relaxation in (001)-oriented InAs/GaSb superlattices, the electron
spin lifetime is expected to be considerably longer in (110)
InAs/GaSb superlattices. Additionally, suppression of the BIA
contribution in (110) heterostructures has been
predicted\cite{DK:1986} and recently verified experimentally in
GaAs/AlGaAs quantum wells.\cite{Ohno:1999} This suppression has
not been seen in any other material system. Here we observe it in
InAs/GaSb superlattices.

We measured spin relaxation times in (110) and (001) InAs/GaSb
superlattices using a mid-infrared femtosecond optical parametric
oscillator and polarization sensitive differential transmission
experiments. We observed an increase in $T_{1}$ by more than an
order of magnitude for (110) growth. Theoretical calculations
using a nonperturbative 14-band ${\bf k\cdot p}$ nanostructure
model agree well with experiment. Due to the suppression of NIA
and BIA contributions to spin relaxation in (110) InAs/GaSb
superlattices, our calculations indicate that $T_{1}$ is highly
sensitive to asymmetry arising from compositional mixing at the
superlattice interfaces, an effect that may limit $T_1$ in (110)
InAs/GaSb superlattices.

The InAs/GaSb superlattices were grown by molecular beam epitaxy
on (001) or (110)-oriented InAs substrates (+/- 0.1 degree) in a
Fisons VG-80 machine equipped with shuttered EPI group-III
evaporators and shuttered EPI valved group V cracker cells. The
superlattice structures contain 2.1 nm thick InAs layers and 3.7
nm thick GaSb layers, with a total of 85 periods.  The (110) and
(001) superlattices were grown in consecutive runs to minimize
differences in extrinsic sample characteristics. X-ray diffraction
measurements indicate high order resolution of the SL peaks, and
features of the RHEED pattern are highly streaked, indicating good
quality superlattice growth.  The room temperature continuous wave
photoluminescence signal from the superlattices was of similar
intensity and line width for the (001) and (110) structures, and
the carrier recombination time, measured by time-resolved
photoluminescence up-conversion experiments, was found to be
similar in the two superlattices (1-2 ns), indicating consistency
of sample quality.

$T_{1}$ times were measured at low temperature (115 K) using pump
probe experiments involving 200 fs, circularly polarized
mid-infrared pulses (4.0-4.6 $\mu$m).  The laser source for these
measurements is an optical parametric oscillator (OPO) that relies
on parametric down conversion in a 1 mm thick periodically poled
LiNbO$_{3}$ (PPLN) crystal.\cite{Ebrahimzadeh:2001} The OPO is
synchronously pumped by a Titanium sapphire oscillator, and
provides 20 to 100 mW of average idler output power over the
spectral tuning range (2.7-4.6 $\mu$m).  The output pulses have a
full-width at half maximum bandwidth of 18 meV.

The confinement-induced splitting of the band edge heavy-hole and
light-hole states is calculated to be $\geq$ 130 meV for the (001)
and (110) InAs/GaSb superlattices considered here. In this case,
circularly polarized idler pulses tuned close to the band gap will
excite only the heavy-hole to conduction band transition, leading
to a 100\% spin-polarized distribution of electrons. The band gap
of the (110) superlattice is slightly lower ($\approx$ 40 meV)
than the corresponding (001) structure. Band structure
calculations for these superlattices suggest that this difference
is due to lower electron and hole confinement energies in the
(110) superlattice in addition to the differing influence of
strain for different directions of superlattice growth. In order
to compare measured spin lifetimes for identical electron kinetic
energies, the OPO was tuned to 35 meV above the band gap for each
structure. The limited tuning range of the OPO, together with the
low band gaps of the InAs/GaSb superlattices studied [240 meV
(110) and 280 meV (001) at 77 K], restricted our experiments to
temperatures between 77 K and 115 K, over which the spin lifetimes
were not found to vary significantly.

The optically-injected spin-polarized electrons generate
absorption bleaching of the associated interband transition
through state filling. By monitoring the transmission of a weaker,
delayed probe pulse that has the same (SCP) or opposite (OCP)
circular polarization as the pump pulse, $T_{1}$ may be extracted.
The differential transmissivity was measured versus the time delay
between the pump and probe pulses using a liquid N$_{2}$-cooled
InSb detector and lock-in detection methods. In our experiments,
care was taken to ensure that the pump beam impinges on the
superlattice sample at normal incidence. In this case, the probe
beam interrogates the relaxation time of electron spins polarized
in the growth direction.  The optically injected carrier density
was estimated using the measured pump pulse fluence and the
calculated absorption spectrum for each structure to be
1$\times$10$^{16}$ cm$^{-3}$ for (110) and 3$\times$10$^{16}$
cm$^{-3}$ for (001), with an uncertainty of $\pm$ 40 \%.

\begin{figure}[htb]
    \includegraphics[width=7.5cm]{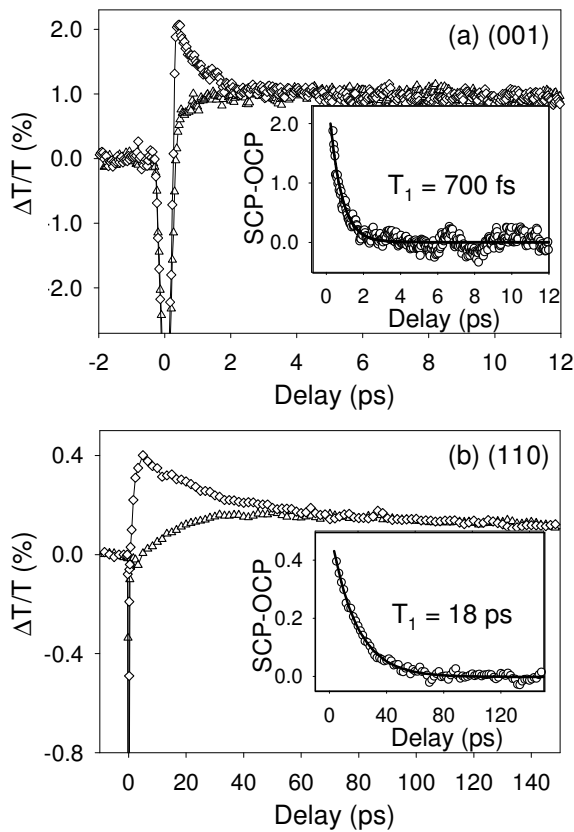}
    \caption{Results of differential transmission experiments on InAs/GaSb superlattices with (a): (001) and (b): (110) directions of
    growth.  Data for conditions of same circular (SCP) and opposite circular (OCP) polarization geometries are indicated by open
    diamonds and triangles, respectively.  Note the time scales on the x-axes in (a) and (b), which differ by more than an
    order of magnitude.  Inset: single exponential fit to the difference between the decay curves for SCP
     and OCP. }
    \label{fig:SpinLifetimes}
\end{figure}

Fig.~\ref{fig:SpinLifetimes} shows the results of
polarization-sensitive differential transmission measurements on
InAs/GaSb superlattices with (001)
(Fig.~\ref{fig:SpinLifetimes}(a)) and (110)
(Fig.~\ref{fig:SpinLifetimes}(b)) directions of growth. Note the
time scales on the x-axes in Fig.~\ref{fig:SpinLifetimes}(a) and
Fig.~\ref{fig:SpinLifetimes}(b), which differ by more than an
    order of magnitude. The large negative
pulse-width-limited feature appearing in both data sets originates
from two-photon absorption in the InAs substrates. This feature
provides a convenient marker for zero time delay, but may
otherwise be ignored for the purpose of this discussion.  The loss
of spin polarization in the optically-injected carrier
distribution is indicated by the convergence of the differential
transmission signals for the SCP and OCP polarization geometries.
>From a comparison of the data in Fig.~\ref{fig:SpinLifetimes}(a)
and Fig.~\ref{fig:SpinLifetimes}(b), it is clear that spin
relaxation occurs on a much longer time scale in the (110)
InAs/GaSb superlattice compared to the (001) structure. The OCP
signal in Fig.~\ref{fig:SpinLifetimes}(b) grows from $\approx 0$,
indicating that the optically excited carriers are initially fully
spin polarized, and that no detectable Coulomb screening effects
are present.  In the data for the (001) superlattice in
Fig.~\ref{fig:SpinLifetimes}(a), the non-zero bleaching signal in
the OCP data immediately following the two-photon absorption
feature is evidence of spin decay during pulse overlap. $T_1$
times are obtained by performing a single exponential fit to the
difference between the SCP and OCP curves. Fits were restricted to
delay values $\geq$ 350 fs to avoid the two-photon absorption
feature. From the data in Fig.~\ref{fig:SpinLifetimes}, we find
$T_{1}$ values of 700 fs and 18 ps for the (001) and (110)
superlattices respectively, indicating that the spin relaxation
time in InAs/GaSb superlattices is strongly enhanced with (110)
growth.

The much longer spin lifetime in the (110) InAs/GaSb superlattice
relative to the corresponding (001) structure is primarily
attributed to the elimination of the NIA contribution to spin
relaxation. Although a larger $T_{1}$ is expected in (110)
superlattices due to suppression of BIA,\cite{DK:1986} for the
short period no-common-atom superlattices investigated here the
more crucial distinction arises from the nature of the interfaces.
Fig.~\ref{fig:Interfaces} contains a schematic diagram of the
interfaces in InAs/GaSb superlattices for conditions of growth
along the (001) [Fig.~\ref{fig:Interfaces}(a)] and (110)
[Fig.~\ref{fig:Interfaces}(b)] directions. For the (001)
superlattice, the potential at the interfaces is asymmetric due to
tetrahedral bonding between the bulk constituents (InAs and GaSb)
that share no atom in common.  This effect is referred to as NIA.
As in the case of BIA, NIA generates a pseudomagnetic field that
serves to relax the electron spins. It has been shown
previously\cite{Olesberg:2001} that NIA strongly dominates the
electron spin decay process in a similar (001)-oriented InAs/GaSb
superlattice grown on a GaSb substrate. The subpicosecond spin
lifetime extracted from the data of
Fig.~\ref{fig:SpinLifetimes}(a) is consistent with these earlier
studies.\cite{Olesberg:2001} In contrast, for an InAs/GaSb
superlattice grown in the (110) direction, NIA is absent because
the interfaces are mixed, containing equal numbers of both anions
and cations (see Fig.~\ref{fig:Interfaces}(b)).  In this case, the
interface potential is symmetric. Since spin decay in the (001)
superlattice is strongly dominated by NIA, with its elimination in
the (110) structure the spin lifetime is expected to sharply
increase, consistent with the results in
Fig.~\ref{fig:SpinLifetimes}(b).

In order to interpret these findings, we have performed
calculations of spin relaxation times due to the precessional
D'yakonov Perel mechanism, which has been found to dominate in
III-V semiconductors and their heterostructures for temperatures
above 77
K.\cite{ALS:book,Olesberg:2001,Lau:2001,OpticalOrientationBook,Kikkawa:1998,Tacheuchi:1997,Terauchi:1999,Boggess:2000}
We employ a nonperturbative ${\bf k\cdot p}$ nanostructure model
solved in a fourteen-band restricted basis set, in which BIA
contributions are included naturally\cite{Lau:2001} and NIA
contributions are introduced as previously
described.\cite{Olesberg:2001} Parameters governing the interface
asymmetry were taken from reference 7. Neutral impurities were
taken as the dominant source of momentum scattering.  The spin
lifetime varies inversely with momentum scattering time, $T_1 \sim
\tau_{p}^{-1}$.  We find good agreement with experiment using
$\tau_p$ = 190 fs, which is within the typical range (100-200 fs)
for the III-V semiconductors.

The calculated spin relaxation time for the (001) superlattice is
$T_{1}$ = 715 fs in good agreement with the experimental value of
700 fs. NIA was found to strongly dominate the electron spin
relaxation process for the (001) structure, consistent with
earlier findings.\cite{Olesberg:2001} For the (110) InAs/GaSb
superlattice, our calculations indicate that, with the elimination
of NIA and the strong suppression of the remaining BIA
contribution,\cite{DK:1986} the spin lifetime is expected to be
$\sim$ 600 ps.  The much shorter $T_{1}$ observed experimentally
(18 ps) indicates that additional contributions to the spin
relaxation process are present and substantially dominate over
BIA.  The electron-hole exchange interaction (the Bir-Aronov-Pikus
mechanism\cite{BAP:1976}(BAP)) is not a favorable candidate to
account for the measured spin lifetime in the (110) superlattice
due to the small electron-hole
overlap\cite{Wagner:1993,Vinattieri:1993} characteristic of these
broken-gap InAs/GaSb superlattices (roughly 20\% compared to
GaAs/AlGaAs quantum wells) and because the timescale of spin decay
(18 ps) is much shorter than that found previously for the BAP
interaction.\cite{Ohno:1999,OpticalOrientationBook} The
Elliott-Yafet mechanism, which increases in importance with
decreasing band gap,\cite{OpticalOrientationBook,Chasalviel:1975}
may limit $T_{1}$ in (110) InAs/GaSb superlattices, although a
detailed theoretical treatment for this mechanism in III-V
semiconductor superlattices is not presently available.

Because of the large spin-orbit interaction and small band gap in
InAs/GaSb superlattices, a substantial contribution to the spin
decay process could arise from a small degree of structural
inversion asymmetry (SIA) associated with compositional mixing at
the superlattice interfaces. In order to assess the sensitivity of
$T_1$ to such effects, the spin lifetime was calculated assuming
that the interface on one side of the InAs layers is not
compositionally abrupt, corresponding to a \emph{single monolayer}
of In$_{0.5}$Ga$_{0.5}$As$_{0.5}$Sb$_{0.5}$.  Such a situation may
arise due to differing growth kinetics at the normal and inverted
interfaces, an effect which is commonly present in InAs/GaSb
superlattices.\cite{Feenstra:1994,Steinshnider:2000,Lew:1997,Wang:1995}
With the inclusion of the mixed composition monolayer, the
calculated spin lifetime in the (110) InAs/GaSb superlattice was
reduced by a factor of 20, yielding a value of 29 ps, in line with
the experimental value of 18 ps. This dramatic reduction in $T_1$
occurs because, although the pseudomagnetic field associated with
BIA is primarily in the growth direction for (110)-oriented
superlattices, SIA introduces a field component in the plane of
the superlattice that serves to relax the growth-direction
polarized electron spins injected in these optical experiments. In
contrast, for the (001)-oriented superlattice including such
compositional mixing has a much smaller influence on the
calculated spin lifetimes because, for (001) superlattices, both
the psuedomagnetic fields associated with BIA\cite{DK:1986} and
NIA have large in-plane components. Detailed STM characterization
of the present (110) superlattice would help in the conclusive
identification of the physical process limiting $T_{1}$.

In summary, we observe an enhancement of the electron spin
relaxation time in InAs/GaSb superlattices by more than an order
of magnitude (25 times) with (110)-oriented growth using
polarization sensitive pump probe techniques and a mid-infrared,
subpicosecond PPLN OPO. This enhancement is attributed to
suppression of the NIA and BIA contributions to the spin decay
process in (110) superlattices.  Our findings have implications
for potential spintronics applications using InAs/GaSb
heterostructures.

This research is supported by the DARPA MDA972-01-C-0002,
DARPA/ARO DAAD19-01-1-0490 and the Natural Sciences and
Engineering Research Council of Canada.

\end{document}